\documentclass[10pt]{article}
\usepackage[dvips]{graphicx}
\begin{document}
\begin{center}
{\LARGE\bf Exponential Potentials and Attractor Solution of
Dilatonic Cosmology}
 \vskip 0.15 in
 $^\dag$Wei Fang$^1$, $^\ddag$H.Q.Lu$^1$, Z.G.Huang$^2$\\
$^1$Department~of~Physics,~Shanghai~University,\\
~Shanghai,~200444,~P.R.China\\
$^2$Department of Mathematics and Physics,
\\Huaihai Institute of Technology, Lianyungang, 222005, P.R.China
\footnotetext{$\dag$ xiaoweifang$_{-}$ren@shu.edu.cn}
\footnotetext{$\ddag$ Alberthq$_{-}$lu@staff.shu.edu.cn}
 \vskip 0.5 in
\centerline{\bf Abstract} \vskip 0.2 in
\begin{minipage}{5.5in} { \hspace*{15pt}\small \par We present
the scalar-tensor gravitational theory with an exponential
potential in which pauli metric is regarded as the physical
space-time metric. We show that it is essentially equivalent to
coupled quintessence(CQ) model. However for baryotropic fluid
being radiation there are in fact no coupling between dilatonic
scalar field and radiation. We present the critical points for
baryotropic fluid and investigate the properties of critical
points when the baryotropic matter is specified to ordinary
matter. It is possible for all the critical points to be
attractors as long as the parameters $\lambda$ and $\beta$ satisfy
certain conditions. To demonstrate the attractor behaviors of
these critical points, We numerically plot the phase plane for
each critical point. Finally with the bound on $\beta$ from the
observation and the fact that our universe is undergoing an
accelerating expansion, we conclude that present accelerating
expansion is not the eventual stage of universe. Moreover, we
numerically describe the evolution of the density parameters
$\Omega$ and the decelerating factor $q$, and computer the present
values of some cosmological parameters, which are consistent with
current observational data.

  {\bf Keywords:} exponential potential;~scalar-tensor;~dark energy;~critical point;~attractor.\\
 {\bf PACS:}98.80.Cq}
\end{minipage}
\end{center}
\section{Introduction}\hspace*{15 pt}The evidences for the
existence of dark energy have been growing in past few years.
Recently the WMAP three result[1] has dramatically shrunk the
allowed volume in the parameter space. It shows that in a spatial
flat universe the combination of WMAP and the Supernova legacy
survey(SNLS) data yields a significant constraint on the equation
of state of the dark energy $w=-0.97^{+0.07}_{-0.09}$. Though the
$\Lambda$CDM model is still an excellent fit to the WMAP data, it
still does not exclude other alternative model for the candidate
of dark energy. Moreover the well-known fine-tuning and
coincidence problems[2] are yet unsolved in cosmological constant
model. This motivates a wide range of theoretic studies to explain
the observation: such as the conventional "quintessence" scalar
field[3]; the k-essence field[4]; quintom model[5]; holograph dark
energy[6]; Born-Infeld scalar or vector field theory[7]; phantom
model[8] and so on. Additionally, some authors attempt to modify
the conventional gravitational theory instead of involving the
exotic matter[9].
\par In past several years the idea that dilaton field of the scalar-tensor gravitational theory as the dark
energy has been proposed and discussed[10]. In our previous
paper[11],we have considered a dilatonic dark energy model, based
on Weyl-scaled induced gravitational theory. In that paper, we
find that when the dilaton field is not gravitational clustered at
small scales, the effort of dilaton can not change the
evolutionary law of baryon density perturbation, and the density
perturbation can grow from $z\sim10^3$ to $z \sim5$, which
guarantees the structure formation. When dilaton energy is very
small compared the matter energy, potential energy of dilaton
field can be neglected. In this case, the solution of cosmological
scale $a$ has been found[12]. In recent work[13], we consider the
dilaton field with positive kinetic energy and with negative
kinetic energy and find that the coupled term between matter and
dilaton can't affect the existence of attractor solutions. In this
paper we will study the attractor properties of the dynamical
system. The potential we choose for investigation is the
exponential form for its important role in higher-order or
higher-dimensional gravitational theories, string theories and
kaluza-klein model(see the references in Ref[14]). Though the
possible cosmological roles of exponential potential have been
investigated elsewhere[15], here we will investigate its
cosmological implies in our dilatonic cosmology. With the
constraint from the observation we conclude that the present
accelerating expansion is not the eventual stage of universe.

\section{Theoretical model from scalar-tensor gravitational theory}\hspace*{15
pt}The action of Jordan-Brans-Dicke theory is:
\begin{equation}S=\int
d^4x\sqrt{-\gamma}[\phi\widetilde{R}-\omega\gamma^{\mu\nu}\frac{\partial_\mu\phi\partial_\nu\phi}{\phi}-\Lambda(\phi)-\widetilde{L}_{fluid}(\psi)]\end{equation}
where $\widetilde{L}_{fluid}$ is the lagrangian density of cosmic
fluid, $\gamma$ is the determinant of the Jordan metric tensor
$\gamma_{\mu\nu}$, $\omega$ is the dimensionless coupling
parameter, $\widetilde{R}$ is the contracted
$\widetilde{R_{\mu\nu}}$. The metric sign convention is
$(-,+,+,+)$. The quantity $\Lambda(\phi)$ is a nontrivial
potential of $\phi$ field. When $\Lambda(\phi)\neq 0$ the Eq.(1)
describes the induced gravity. $\widetilde{\rho}$ and
$\widetilde{p}$ is respectively the density and pressure of cosmic
fluid. However it is often useful to write the action in terms of
the conformally related Einstein metric. We introduce the dilaton
field $\sigma$ and conformal transformation as follows:
\begin{equation}\phi=\frac{1}{2\kappa^2}e^{\alpha\sigma}\end{equation}
\begin{equation}\gamma_{\mu\nu}=e^{-\alpha\sigma}g_{\mu\nu}\end{equation}
where $\alpha^2=\frac{\kappa^2}{2\omega+3}$, $\kappa^2=8\pi G$.
Using Eqs.(2,3), the action (1) becomes
\begin{equation}S=\int{d^4x\sqrt{-g}[\frac{1}{2\kappa^2}(R(g_{\mu\nu})+\frac{1}{2}g^{\mu\nu}\partial_\mu\sigma\partial_\nu\sigma-W(\sigma))+L_{fluid}(\psi)}]\end{equation}
where
$L_{fluid}(\psi)=\frac{1}{2}g^{\mu\nu}e^{-\alpha\sigma}\partial_\mu\psi\partial_\nu\psi-e^{-2\alpha\sigma}V(\psi)$
\par The transformation Eq.(2) and Eq.(3) are well defined for some
$\omega$ as $-\frac{3}{2}<\omega<\infty$. The conventional
Einstein gravity limit occurs as $\sigma\rightarrow 0$ for an
arbitrary $\omega$ or $\omega\rightarrow\infty$ with an arbitrary
$\sigma$.
\par The nontrivial potential of the $\sigma$ field,
$W(\sigma)$ can be a metric scale form of $\Lambda(\phi)$.
Otherwise, one can start from Eq.(4), and define $W(\sigma)$ as an
arbitrary nontrivial potential. $g_{\mu\nu}$ is the pauli metric.
Cho and Damour et.al pointed out that the pauli metric can
represent the massless spin-two graviton in scalar-tensor
gravitational theory[16]. Cho also pointed out that in the
compactification of Kaluza-Klein theory, the physical metric must
be identified as the pauli metric because of the the wrong sign of
the kinetic energy term of the scalar field in the Jordan frame.
The dilaton field appears in string theory naturally.\par By
varying the action Eq.(4), one can obtain the field equations of
Weyl-scaled scalar-tensor gravitational theory.
$$R_{\mu\nu}-\frac{1}{2}g_{\mu\nu}R=-\frac{\kappa^2}{3}\{[\partial_\mu\sigma\partial_\nu\sigma-\frac{1}{2}g_{\mu\nu}\partial_\rho\sigma\partial^\rho\sigma]
-g_{\mu\nu}W(\sigma)$$
\begin{equation}+e^{-\alpha\sigma}[\partial_\mu\psi\partial_\nu\psi-\frac{1}{2}g_{\mu\nu}\partial_\rho\psi\partial^\rho\psi]
-g_{\mu\nu}e^{-2\alpha\sigma}V(\psi)\}\end{equation}
\begin{equation}\bigtriangleup\sigma=\frac{dW(\sigma)}{d\sigma}-\frac{\alpha}{2}e^{-2\alpha\sigma}g^{\mu\nu}\partial_\mu\psi\partial_\nu\psi
-2\alpha e^{-2\alpha\sigma}V(\psi)\end{equation}
\begin{equation}\bigtriangleup\psi=-\alpha g_{\mu\nu}\partial_\mu\psi\partial_\nu\sigma+e^{-\alpha\sigma}\frac{dV(\psi)}{d\psi}\end{equation}
where "$\bigtriangleup$" denotes the D'Alembertian. We assume that
the energy-momentum tensor $T_{\mu\nu}$ of cosmic fluid is
\begin{equation}T_{\mu\nu}=(\rho+p)U_\mu U_\nu+pg_{\mu\nu}\end{equation}
where the density of energy
\begin{equation}\rho=\frac{1}{2}\dot{\psi}^2+e^{-\alpha\sigma}V(\psi)\end{equation}
the pressure
\begin{equation}p=\frac{1}{2}\dot{\psi}^2-e^{-\alpha\sigma}V(\psi)\end{equation}
\par In  FRW metric $ ds^2=-dt^2+a^2(t)(dx^2+dy^2+dz^2)$, we can obtain
following equations from Eqs.(5,6,7):
\begin{equation}(\frac{\dot{a}}{a})^2=\frac{\kappa^2}{3}(\frac{1}{2}\dot{\sigma}^2+W(\sigma)+e^{-\alpha\sigma}\rho)\end{equation}
\begin{equation}\ddot{\sigma}+3H\dot{\sigma}+\frac{dW}{d\sigma}=\frac{1}{2}\alpha
e^{-\alpha\sigma}(\rho-3p)\end{equation}
\begin{equation}\dot{\rho}+3H(\rho+p)=\frac{1}{2}\alpha\dot{\sigma}(\rho+3p)\end{equation}
\begin{equation}\dot{H}=-\frac{\kappa^2}{2}[\dot\sigma^2+e^{-\alpha\sigma}(\rho+3p)]\end{equation}

\par where we specify the cosmic fluid $\rho$ as the baryotropic
matter $\rho_b$ with a equation of state $p_b=w_b\rho_b$. $w_b=0$
for ordinary matter, $w_b=\frac{1}{3}$ for radiation, $W(\sigma)$
is exponentially dependent on $\sigma$ as
$W(\sigma)=W_0e^{-\lambda\kappa\sigma}$. The effective energy
density of dilaton scalar field is
$\rho_{\sigma}=\frac{1}{2}\dot{\sigma}^2+W(\sigma)$, the effective
pressure of dilaton scalar field is
$p_{\sigma}=\frac{1}{2}\dot{\sigma}^2-W(\sigma)$ and
$p_{\sigma}=w_{\sigma}\rho_{\sigma}$. Here we should note that in
Weyl-scaled scalar-tensor gravitational theory the coupling of the
baryotropic matter with dilaton field $\sigma$ is more natural.
\par We can get the solution of Eq.(13) for the density of baryotropic
energy:
\begin{equation}\rho_b\propto e^{\frac{1}{2}\alpha(1+3w_b)\sigma}a^{-3(1+w_b)}\end{equation}
For matter, $w_m=0$, $\rho_m\propto
e^{\frac{1}{2}\alpha\sigma}a^{-3}$, and for radiation,
$w_r=\frac{1}{3}$, $\rho_r\propto e^{\alpha\sigma}a^{-4}$
\section{Critical points and the attractor solution}\hspace*{15
pt}In this section, we investigate the global structure of the
dynamical system via a phase plane analysis.
\par We define
\begin{equation}x=\frac{\kappa\dot\sigma}{\sqrt6 H}, y=\frac{\kappa\sqrt{W(\sigma)}}{\sqrt3 H},
 z=\frac{\kappa\sqrt{e^{-\alpha\sigma}\rho_b}}{\sqrt3H}\end{equation} Then Eqs.(12-14) can be written as a
plane autonomous system:
\begin{equation}\frac{dx}{dN}=\frac{\sqrt6\alpha}{4\kappa}(1-3w_b)z^2-3x+\frac{\sqrt6\lambda}{2}y^2+3x^3+\frac{3}{2}(1+w_b)xz^2\end{equation}
\begin{equation}\frac{dy}{dN}=-\frac{\sqrt6\lambda}{2}xy+3x^2y+\frac{3}{2}(1+w_b)yz^2\end{equation}
\begin{equation}\frac{dz}{dN}=-\frac{\sqrt6\alpha}{2\kappa}xz+\frac{\sqrt6\alpha}{4\kappa}(1+3w_b)xz-\frac{3}{2}(1+w_b)z+3x^2z+\frac{3}{2}(1+w_b)z^3\end{equation}
where $N=ln(a)$, the constraint Eq.(11) becomes:
\begin{equation}x^2+y^2+z^2=1\end{equation}
the density parameter of $\sigma$ field is
\begin{equation}\Omega_{\sigma}=x^2+y^2\end{equation}
the equation of state of the dilaton field $\sigma$ is:
\begin{equation}w_{\sigma}=\frac{x^2-y^2}{x^2+y^2}\end{equation}
 In order to investigate the expansive behavior of scale factor $a$, we also represent the
decelerating factor:
\begin{equation}q=-\frac{\ddot{a}a}{\dot{a}^2}=
\frac{3}{2}[(1-w_b)x^2-(1+w_b)y^2+(w_b+\frac{1}{3})]\end{equation}
Using Eq.(20), we rewrite Eqs.(17-19) as follows:
\begin{equation}\frac{dx}{dN}=-3x+\lambda\sqrt{\frac{3}{2}}y^2+\frac{3}{2}x[2x^2+(1+w_b)(1-x^2-y^2)]+\frac{\sqrt6}{4}\beta(1-3w_b)(1-x^2-y^2)\end{equation}
\begin{equation}\frac{dy}{dN}=y\left(-\sqrt{\frac{3}{2}}\lambda x+\frac{3}{2}[2x^2+(1+w_b)(1-x^2-y^2)]\right)\end{equation}
where $\beta=\alpha/\kappa$. It is more convenient to investigate
the global properties of the dynamical system Eqs.(24,25) than
Eqs.(17-19). Because the case $y<0$ corresponds to contracting
universes, we will only consider the case $y \geq 0$ in the
following discussion.
\par We can generally find five fixed points(critical points) where
$dx/dN$ and $dy/dN$ both equal $0$(TABLE 1) :
\begin{center}
\begin{tabular}{|c|c|c|c|c|c|}
\hline
Points & P1 & P2 & P3 & P4 & P5\\
\hline
$x$ & $\frac{\sqrt6\beta(1-3w_b)}{6(1-w_b)}$ & 1 & -1 & $\frac{\sqrt6}{6}\lambda $& $\frac{\sqrt6(1+w_b)}{2\lambda-(1-3w_b)\beta}$\\
\hline $y$ & 0 & 0 & 0 & $(1-\frac{\lambda^2}{6})^{1/2}$ &
$\frac{\sqrt{\beta^2(3w_b-1)^2+2\lambda\beta(3w_b-1)-6({w_b}^2-1)
}}{2\lambda-(1-3w_b)\beta}$\\
\hline
\end{tabular}
\end{center}

\par Depending on the different values of $w_b$, $\lambda$ and
$\beta$, we will study the stability of these critical points.
Before investigating the stability of critical points, we should
point out that for radiation $w_b=\frac{1}{3}$,
$\rho_r=e^{\alpha\sigma}a^{-4}$, there are no coupling between
dilaton field $\sigma$ and radiation in Eq.(11). Since
$\rho_{\sigma}=3p_{\sigma}$, the right of Eq.(12) equals zero and
therefore the radiation density does not contribute to Eq.(12). In
this case, the form of Eqs.(11,12,24,25) are very similar with
conventional scalar field and the corresponding stability of the
critical points are presented in[14]. Therefore we only pay
attention to the case $w_b=0$(matter). In this case, we will find
that the properties of critical points are more distinctive than
the case $w_b=\frac{1}{3}$. We present the critical points and its
properties in following TABLE 2:
\begin{center}
\begin{tabular}{|c|c|c|c|c|c|c|}
\hline
Points&$x$& $y$ & Stability & $\Omega_{\sigma}$ & $\omega_{\sigma}$ & $q$ \\
\hline
 P1&$\frac{\sqrt6}{6}\beta$ & $0$ &
$\lambda_-<\beta<\sqrt6$($\lambda>\sqrt6$)  & $\frac{\beta^2}{6}$ & 1 & $\frac{\beta^2+2}{4}$ \\
& & & or $-\sqrt6<\beta<\lambda_+$($\lambda<-\sqrt6$)& & & \\
 \hline
P2&$1$ & $0$ &  $\lambda>\sqrt6$ and $\beta>\sqrt6$ & $1$ & $1$ & $2$ \\
\hline
P3&$-1$ & $0$ & $\lambda<-\sqrt6$ and $\beta<-\sqrt6$ & $1$ & $1$ & $2$ \\
\hline
P4&$\frac{\sqrt6}{6}\lambda$&$(1-\frac{\lambda^2}{6})^{1/2}$&$\lambda^2<6$ and $\beta^2<24$& $1$ & $\frac{\lambda^2}{3}-1$ & $\frac{\lambda^2}{2}-1$ \\
\hline
P5&$\frac{\sqrt6}{6}\lambda$&$(1-\frac{\lambda^2}{6})^{1/2}$&$max(\beta_-,-\sqrt6)<$& $1$ & $\frac{\lambda^2}{3}-1$ & $\frac{\lambda^2}{2}-1$ \\
& & & $\lambda<min(\beta_+,\sqrt6))$& & & \\
 \hline
P6&$\frac{\sqrt6}{2\lambda-\beta}$& $[\frac{\beta^2-2\lambda\beta+6}{(2\lambda-\beta)^2}]^{1/2}$ & see Eq.(31)&$ A$& $B $&$ C$ \\
\hline
\end{tabular}
\end{center}
\par where $A=\frac{\beta^2-2\lambda\beta+12}{(2\lambda-\beta)^2}$,
$B=-1+\frac{12}{\beta^2-2\lambda\beta+12}$ and
$C=\frac{(2\lambda-\beta)(\lambda+\beta)}{(2\lambda-\beta)^2}$. We
define $\lambda_{\pm}=\lambda\pm\sqrt{\lambda^2-6}$,
$\beta_{\pm}=\frac{\beta\pm\sqrt{\beta^2+48}}{4}$.
\par From Table 2, we know that for every critical point, they
are all possible to be stable, but for every value of the
parameters $\lambda$ and $\beta$, there is one and only one stable
critical point; the others are unstable.
\par P1 is a stable point(Fig1), which corresponds to a late-time decelerating
rolling attractor solution where neither the dilaton nor the
baryotropic fluid entirely dominates the evolution. Note that
$\Omega_{\sigma}=\frac{\beta^2}{6}$, which is irrelevant to the
potential parameter $\lambda$.
\par The properties of P2(Fig2) and P3(Fig3)
are very similar, they both correspond to a late-time decelerating
attractor solutions. In this case the right of Eq.(11) is
dominated by the kinetic energy of the dilaton field, which are
different with the critical point P1. All the three points
describe the dilaton field behaving as a "stiff" matter with a
equation of state $w_{\sigma}=1$. These three points are obtained
from $y=0$ in Eq.(25).
\par However if \begin{equation} -\sqrt{\frac{3}{2}}\lambda x+\frac{3}{2}[2x^2+(1+w_b)(1-x^2-y^2)]=0\end{equation}
we can obtain other critical points.
 From Eqs.(24,25)($dx/dN=0,dy/dN=0$) we get following
 equation:
  \begin{equation}(\beta-2\lambda)x^2-\frac{\sqrt6}{6}[\lambda(\beta-2\lambda)-6]x-\lambda=0\end{equation}
   Solving Eq.(27), we can obtain critical points P4-P6.
    there are two different cases:
   \par \textbf{Case 1} If $\beta=2\lambda$, there exists only one solution(P4, Fig4) for
   Eq.(27). It is a stable node point for
$|\lambda|<\sqrt6$ and corresponds to an accelerating expansive
universe for $|\lambda|<\sqrt2$.
\par \textbf{Case 2} If $\beta\neq 2\lambda$, there are two solutions for
Eq.(27) and therefore exist two critical points(P5, P6).
\par For critical point P5 it is stable for $max(\beta_-, -\sqrt6)<\lambda<min(\beta_+,\sqrt6)$.
 P5 corresponds to a
late-time attractor solution(Fig5) where the density of dilaton
field will dominate the universe. The universe will undergo a
stage of accelerating expansion if $|\lambda|<\sqrt2$.
\par There are two constraints on the parameters $\beta$ and $\lambda$ for point P6. Since
$y_5=[\frac{\beta^2-2\lambda\beta+6}{(2\lambda-\beta)^2}]^{1/2}$,
we have the first constraint:
\begin{equation}\beta^2-2\lambda\beta+6\geq 0\end{equation}
The second constraint is from that:
$\Omega_{b}=1-\Omega_{\sigma}\geq0$. So we have:
\begin{equation}2\lambda^2-\lambda\beta-6\geq 0\end{equation}
To be a stable point for P6(Fig6), the parameters $\lambda$ and
$\beta$ have to satisfy another constraint(from the analysis of
the stability of Eqs.(24,25)):
\begin{equation}(2\lambda-\beta)[\lambda-\frac{1}{2}(\beta^2-\beta^3)^{1/3}-\frac{1}{2}\beta]>0\end{equation}
From Eqs.(28-30) we have following constraints on $\lambda$ and
$\beta$:
\begin{equation}
\begin{array}{lll}\lambda\geq\beta_+ & for&
\beta<0;\\
\lambda\leq\beta_-& for& \beta>0;\\
\beta_-\geq\lambda\geq\frac{3}{\beta}+\frac{\beta}{2}
&for & 0>\beta\geq-\sqrt6;\\
\frac{3}{\beta}+\frac{\beta}{2}\geq\lambda\geq\beta_+ &for&
\sqrt6\geq\beta>0;\end{array}\end{equation} Eq.(31) is the
condition for P5 being a stable point.
\par If the parameters $\lambda$ and $\beta$
also satisfy another relation:
\begin{equation}(2\lambda-\beta)(\lambda+\beta)< 0\end{equation} the stable point will correspond to a late-time accelerating
expansive universe where the dark energy can not dominate the
universe entirely. From Eqs.(31,32), we get the range of $\lambda$
and $\beta$ for an accelerating expansive universe:
\begin{equation}
\begin{array}{lll}-\beta>\lambda\geq\beta_+
&for&
\beta<-\sqrt2\\
\beta_-\geq\lambda\geq-\beta &for&
\beta>\sqrt2\end{array}\end{equation}
\par However if
$2\lambda^2-\lambda\beta-6=0$( $\Delta=0$ in Eq.(27)), the two
different critical points(P5 and P6) become one same point. In
this case we find that the largest eigenvalue for linear
perturbations vanishes and we must consider higher-order
perturbations about the critical point to determine its stability.
\par In order to clearly understand the attractor properties of these six critical points,
we plot their phase planes when the values of $\lambda$ and
$\beta$ lie in the corresponding range presented in Table 2.
\vskip 0.3 in
\begin{minipage}{0.45\textwidth}
\includegraphics[scale=0.3,origin=c,angle=270]{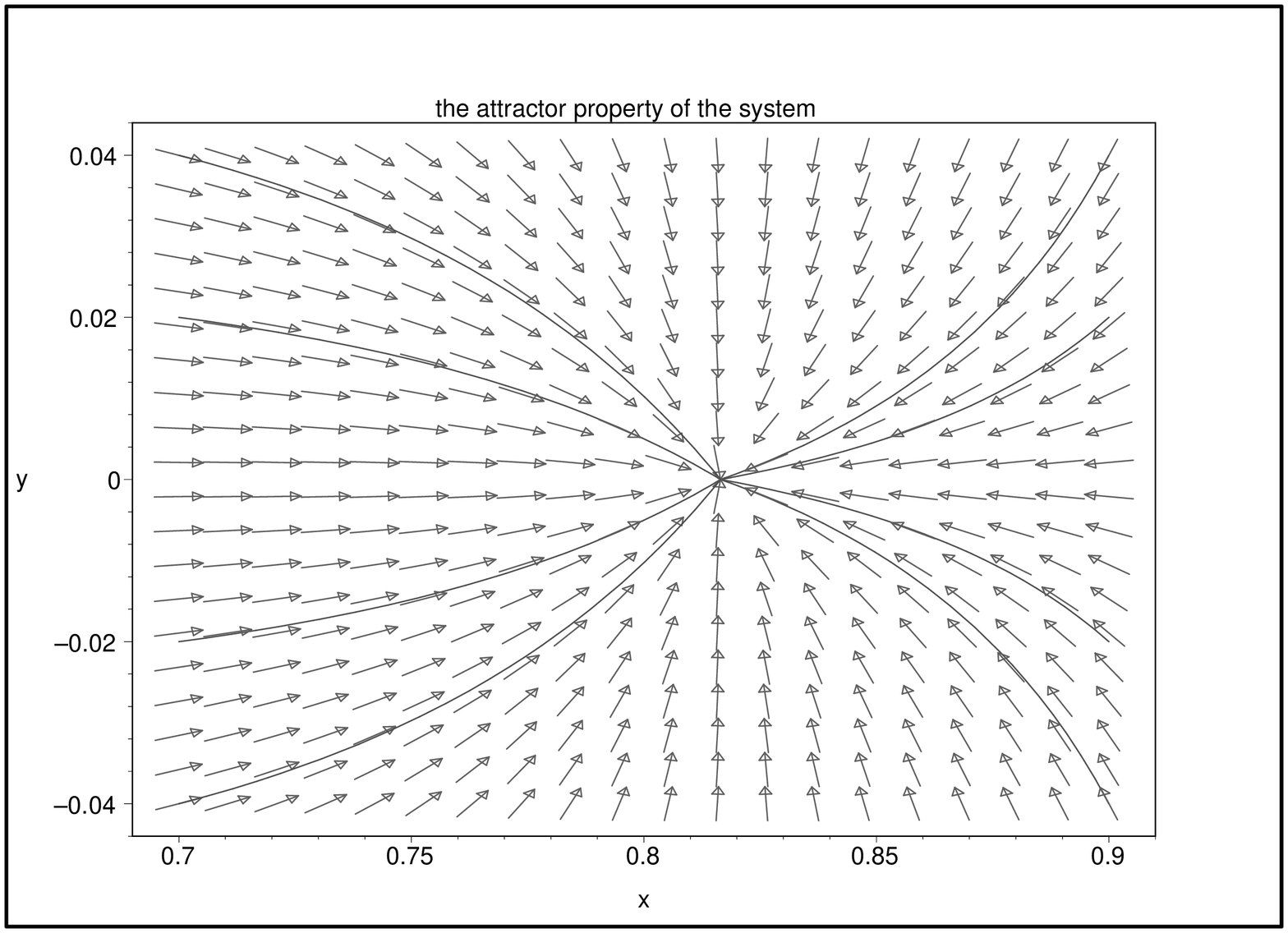}
 {\small  ~Fig1. The phase plane for $\lambda=3$, $\beta=2$(P1).
The late-time attractor is rolling solution with $x_1=\sqrt6/3$,
$y_1=0$,where $\Omega_{\sigma}=2/3$, $q=3/2$ and $w_{\sigma}=1$\\}
\end{minipage}
\hspace{0.1\textwidth}
\begin{minipage}{0.45\textwidth}
\includegraphics[scale=0.3,origin=c,angle=270]{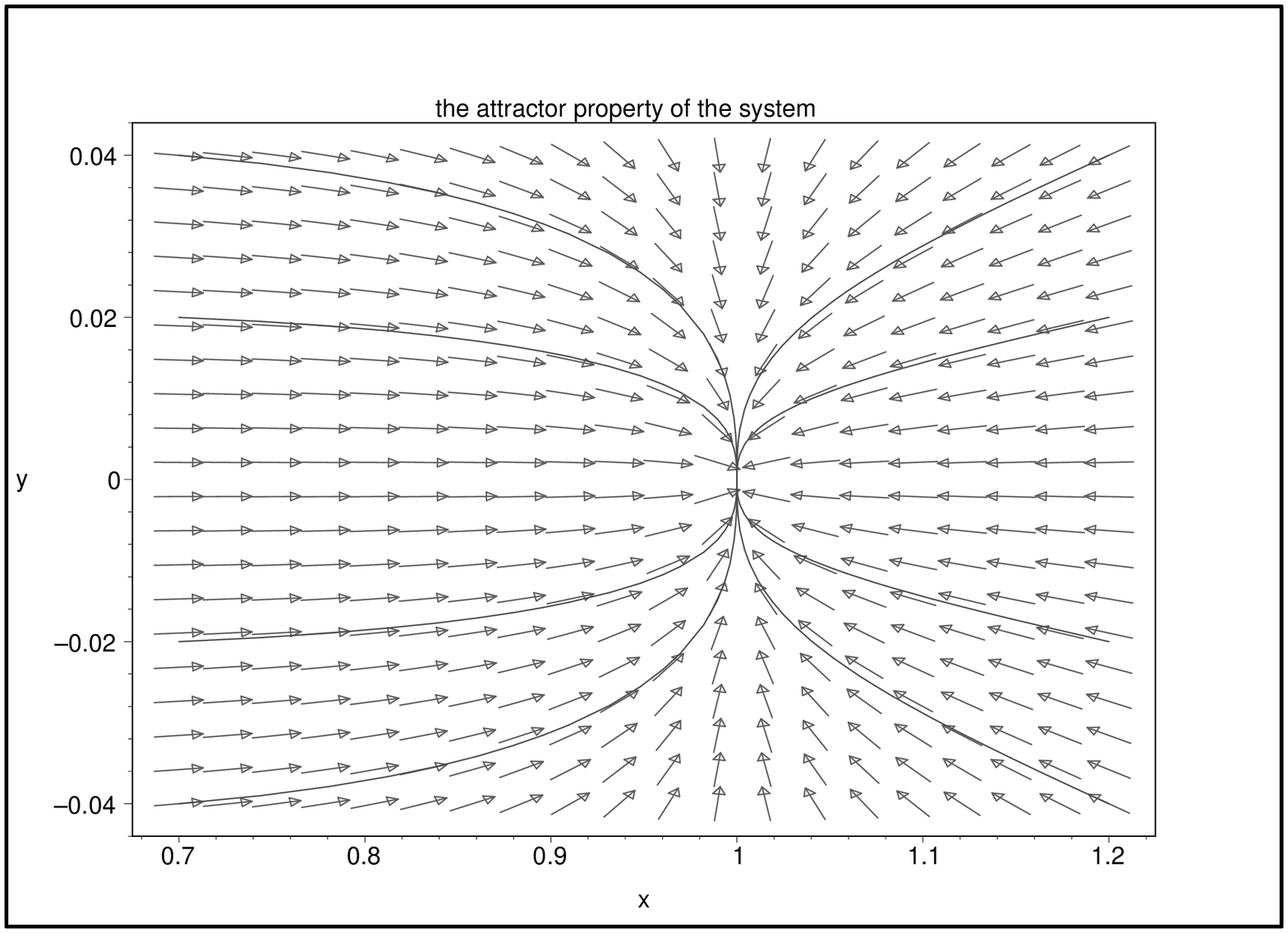}
{\small  ~Fig2. The phase plane for $\lambda=3$, $\beta=4$(P2).
The late-time attractor is a dilaton field dominated solution with
$x_2=1$, $y_2=0$, where $\Omega_{\sigma}=1$, $q=2$ and
$w_{\sigma}=1$\\}
\end{minipage}
~\begin{minipage}{0.46\textwidth}
\includegraphics[scale=0.3,origin=c,angle=270]{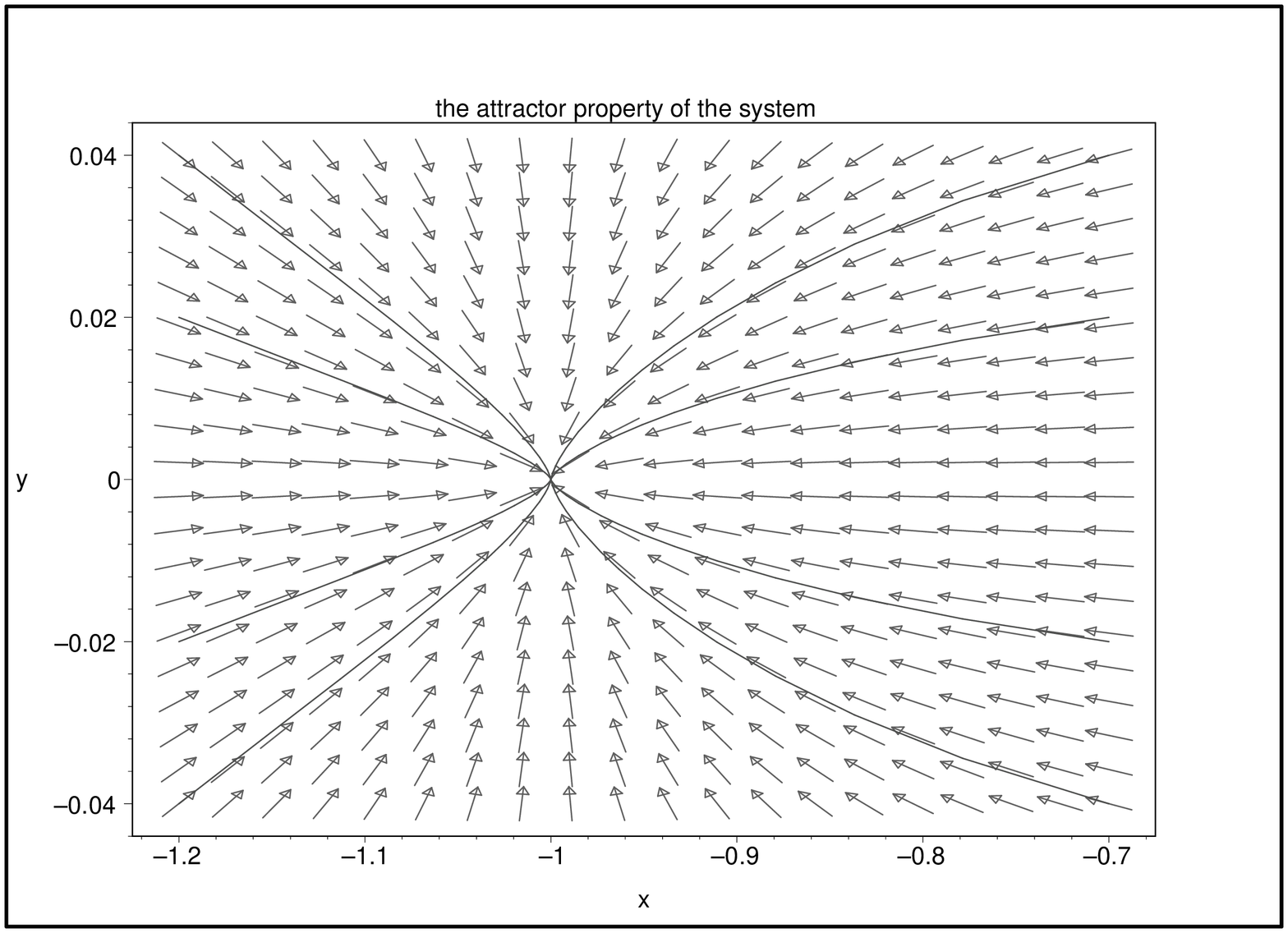}
{\small  ~Fig3. The phase plane for $\lambda=-5$, $\beta=-6$(P3).
The late-time attractor is a dilaton field dominated solution with
$x_3=-1$, $y_3=0$, where $\Omega_{\sigma}=1$, $q=2$ and
$w_{\sigma}=1$\\}
\end{minipage}
\hspace{0.1\textwidth}
\begin{minipage}{0.46\textwidth}
\includegraphics[scale=0.3,origin=c,angle=270]{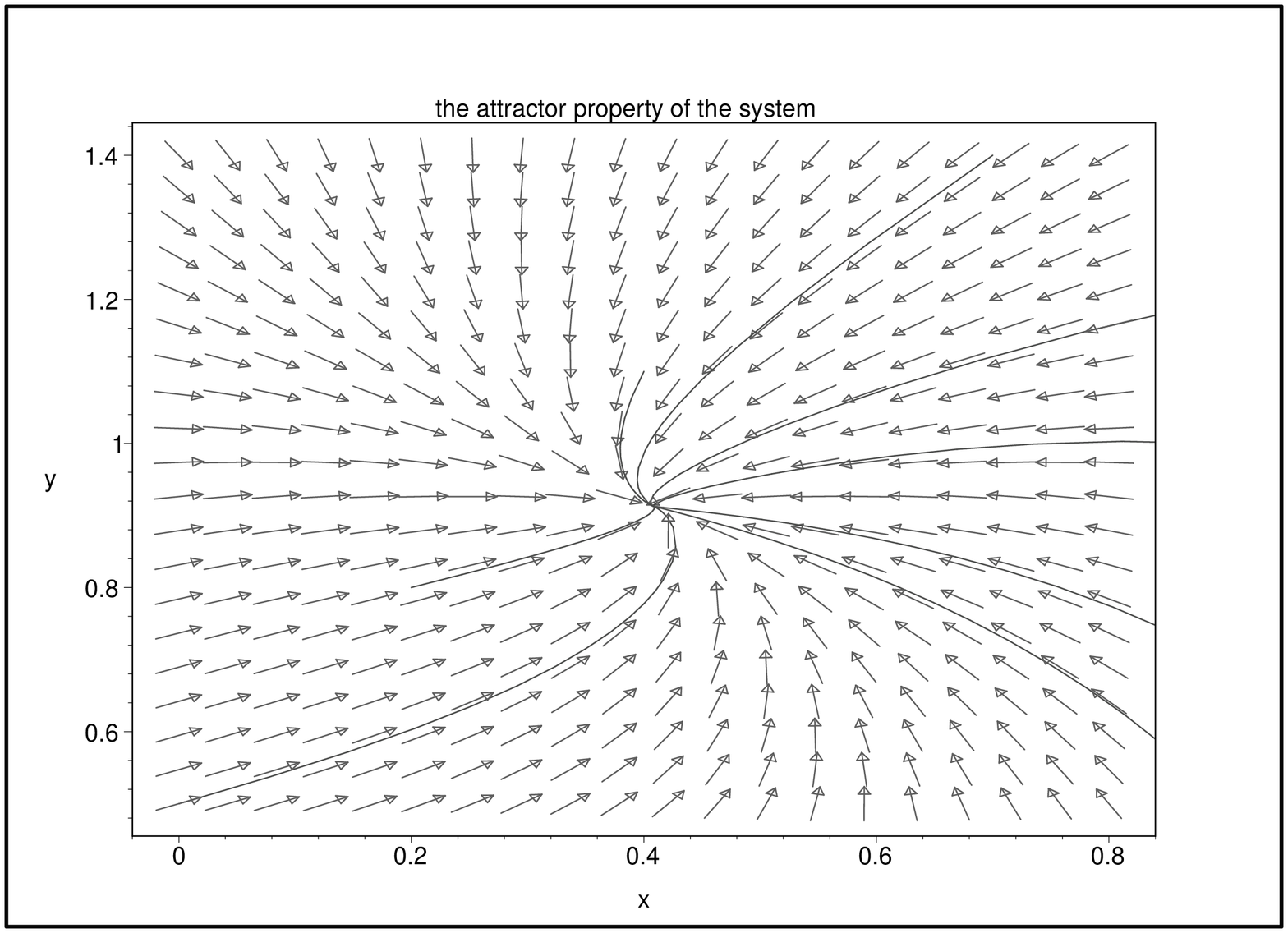}
{\small  ~Fig4.The phase plane for $\beta=2\lambda=2$(P4). The
late-time attractor is a dilaton field dominated solution with
$x_4=\sqrt{1/6}$, $y_4=\sqrt{5/6}$, where $\Omega_{\sigma}=1$,
$q=-1/2$ and $w_{\sigma}=-2/3$\\}
\end{minipage}
\hfill
\begin{minipage}{0.46\textwidth}
\includegraphics[scale=0.3,origin=c,angle=270]{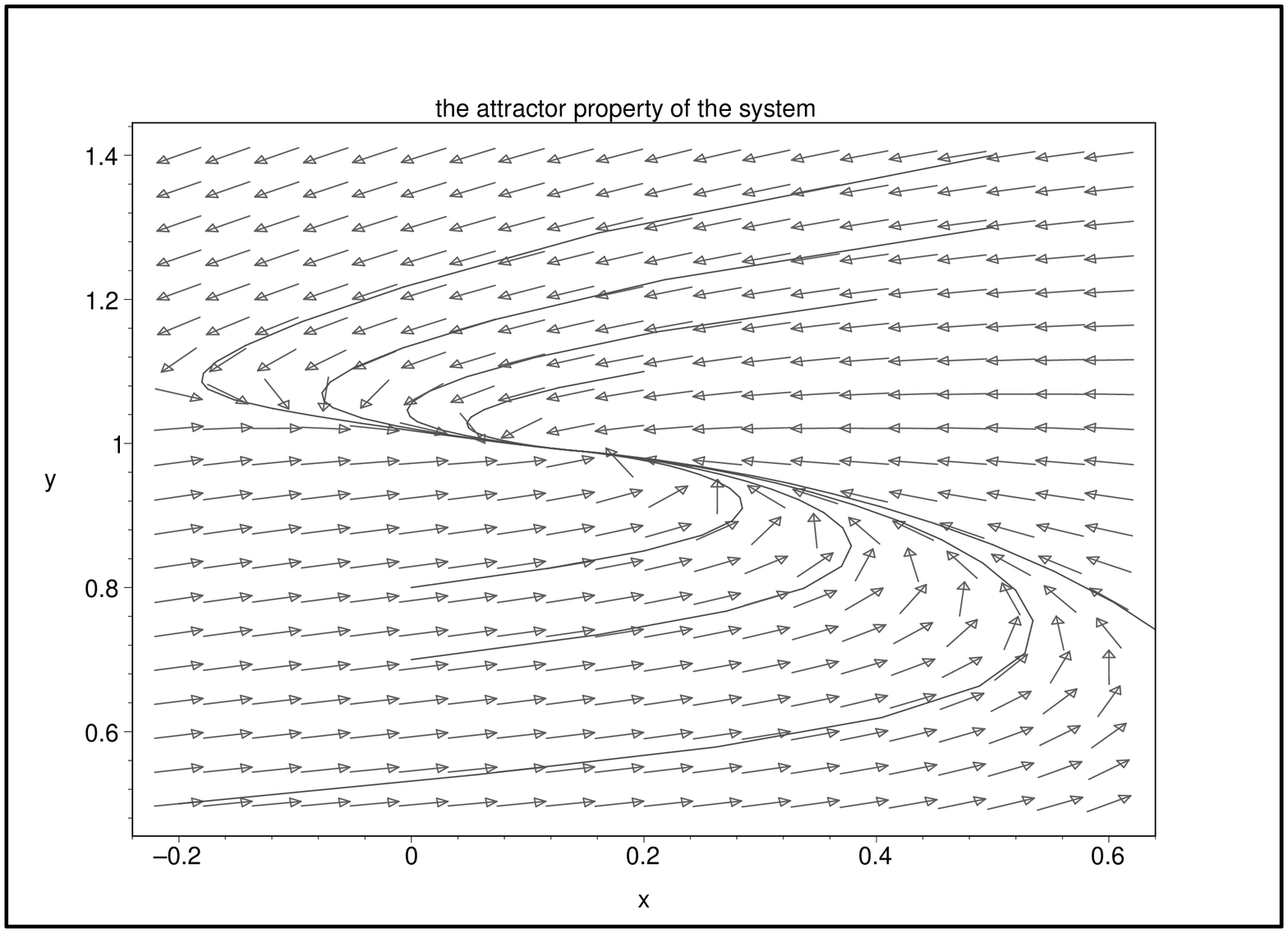}
{\small  ~Fig5.The phase plane for $\lambda=1/3$, $\beta=8$(P5).
The late-time attractor is a dilaton field dominated solution with
$x_5=\sqrt{1/54}$, $y_5=\sqrt{53/54}$, where $\Omega_{\sigma}=1$,
$q\approx-0.944$ and $w_{\sigma}=-0.963$\\}
\end{minipage}
\hfill
\begin{minipage}{0.46\textwidth}
\includegraphics[scale=0.3,origin=c,angle=270]{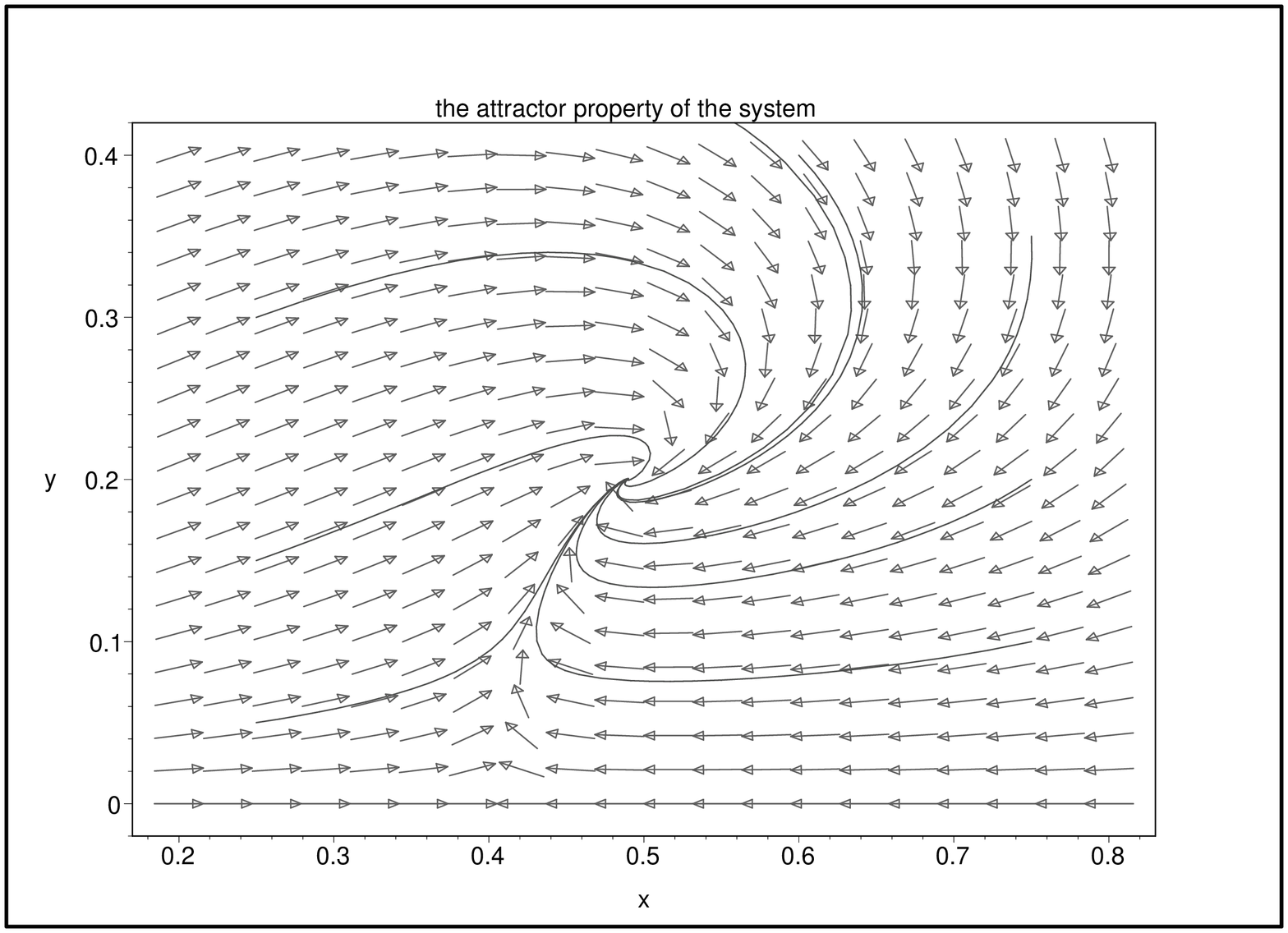}
{\small  ~Fig6.The phase plane for $\lambda=3$, $\beta=1$(P6,
decelerating expansion). The late-time attractor is scaling
solution with $x_6=\sqrt6/5$, $y_6=1/5$, where
$\Omega_{\sigma}=7/25$, $q=4/5$ and $w_{\sigma}=5/7$\\}
\end{minipage}
\hfill

\section{Constraint from observation} \hspace*{15 pt} Firstly we consider the observational
data that the present solar-system gravitational experiments set
(at the $1\sigma$ confidence level) a tight upper bound on
$\beta_{solar-system}^2<10^{-3}$[17]. So we have:
\begin{equation}\beta^2<10^{-3}\end{equation} Therefore with the constraint on $\beta$
and the attractor conditions presented in Table 2, we know that
there only exists four possible late-time attractors(P1, P4, P5,
P6), corresponding to four different cosmological destinies:
\par \textbf{A.} the critical point P1, it is a stable node for
$\lambda_-<\beta<\sqrt6(\lambda>\sqrt6)$ or
$-\sqrt6<\beta<\lambda_+(\lambda<-\sqrt6)$. Considering the upper
bound Eq.(34) on $\beta$, it requires $|\lambda|> 94.88$. In this
case, the universe will be nearly entirely dominated by ordinary
matter($\Omega_{\sigma}=\frac{\beta^2}{6}<\frac{1}{6000}$) and end
with a decelerating expansion. Therefore the present accelerating
expansion is only a transient regime.
\par \textbf{B.} the critical point P4. in this case $\beta=2\lambda$, so from
Eq.(34),we have $\lambda^2<\frac{1}{4}\times 10^{-3}$. In this
case the universe will be entirely dominated by dilatonic scalar
field($\Omega_{\sigma}=1$)and it is undergoing an accelerating
expansion with the value of $ w_{\sigma}\approx -1$ and $q\approx
-1 $. Since the current density parameter of dark energy
$\Omega_{DE}\approx 0.7$, the current cosmological stage is not a
final evolutive stage yet.
\par \textbf{C.} the critical point P5. It is stable for
$max(\beta_-,-\sqrt6)<\lambda<min(\beta_+,\sqrt6))$. From Eq.(34),
the constraint for $\lambda$ is $1.73>\lambda>-1.73$. In this case
the universe will be entirely dominated by dilatonic scalar
field($\Omega_{\sigma}=1$). Whether the universe finally undergoes
an accelerating expansion is determined by the value of
$\lambda$($q=\frac{1}{2}(\lambda^2-2))$. Clearly the current
cosmological stage is not a final evolutive stage yet.
\par \textbf{D.} the critical point P6.
From Eq.(33) we know that there is no accelerating expansion if
$\beta$ satisfies Eq.(34), so the current stage is still not a
final evolution in this case.
\par The second constraint is from the fact that our universe is now undergoing an
accelerating expansion and is dominated by the dark energy. The
contribution of dark energy is $\Omega_{\sigma}\approx0.7$ and its
state parameter($\omega_{\sigma}$) is very close to $-1$. So, we
can exclude the points P1 and P6, which do not admit accelerating
expansion. Therefore there only exist two critical points P4 and
P5 which satisfy above constraints. So we can conclude that the
universe will finally be completely dominated by dilatonic scalar
field, hence the current cosmological state is not our ultimate
destiny and we are all on the way toward final late-time
attractor. Furthermore, We plot the evolution of density parameter
$\Omega_{\sigma}$, $\Omega_{m}$ and $\Omega_{r}$(Fig7, where we
consider the baryotropic matter $\rho_b$ as matter and radiation,
not only one). Fig8 is the evolution of the decelerating factor
$q$ with respect to $N$, which quantitatively indicates that the
universe is speeding up at present but underwent a decelerated
regime at the recent past. Since
$T_{eq}\simeq5.64(\Omega_0h^2)eV\simeq2.843\times10^4K$(where we
set $h=0.71$), $T_0\simeq2.7K$, $a_i=a_{eq}=1$, the scale factor
at the present epoch $a_0$ would be nearly $1.053\times10^4$, then
we know at present epoch $N_0=lna_0\simeq9.262$. We calculate the
present value of $\Omega_{\phi_0}\simeq0.701$, decelerating factor
$q_0\simeq-1.085$ and the transition redshift $Z_T\simeq0.670$ in
our model. It is consistent with current observation data. \vskip
0.3 in
\begin{minipage}{0.46\textwidth}
\includegraphics[scale=0.92,origin=c,angle=0]{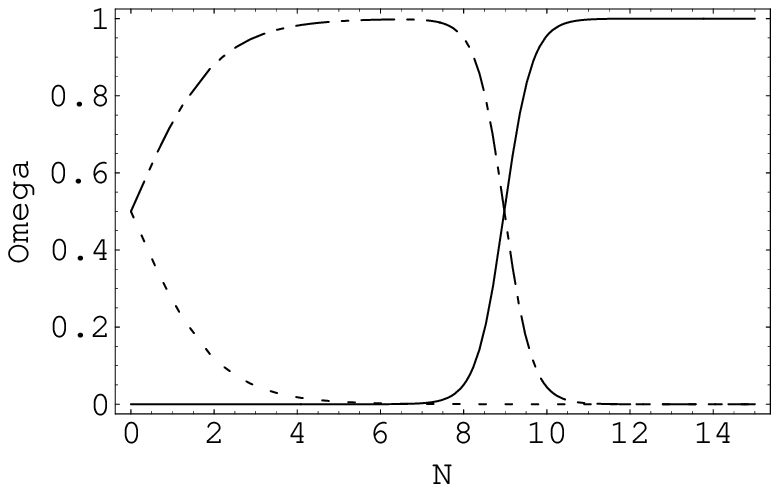}
{\small  ~Fig7.The evolution of density parameter $\Omega$ with
respect to $N$. $\lambda=0.01$, $\beta=-0.02$. Solid line is for
the dilaton field $\Omega_{\sigma}$, dotted line is for radiation
$\Omega_{r}$ and dashed line is for matter $\Omega_{m}$.
\\}
\end{minipage}
\hfill
\begin{minipage}{0.46\textwidth}
\includegraphics[scale=0.92,origin=c,angle=0]{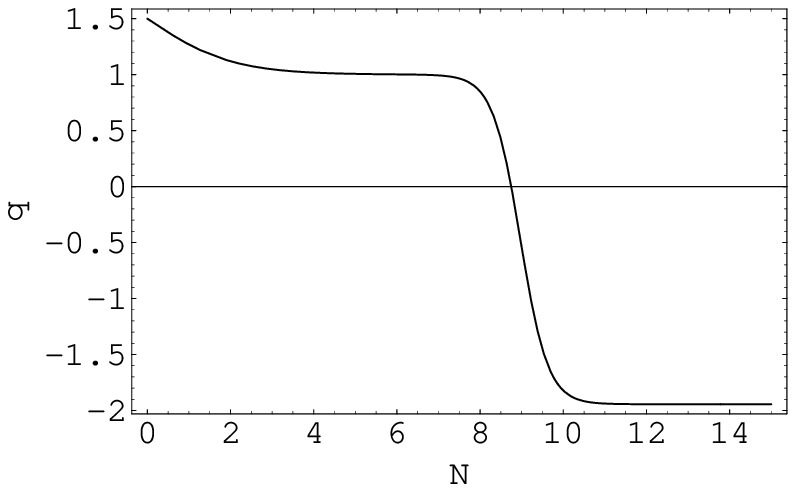}
{\small  ~Fig8.The evolution of decelerating factor $q$ with
respect to $N$. It clearly shows that the universe underwent a
decelerated regime at the recent past with the transition redshift
$Z_T\simeq0.670$
\\}
\end{minipage}

 \hspace*{15 pt}
 \section{Summary}In this paper, we studied the global properties of
dilaton universe in the presence of dilatonic scalar field
$\sigma$ coupling to a baryotropic perfect fluid. We find the
general solution( Eq.(15)) for the density of baryotropic energy.
We show that if the baryotropic matter is radiation($w_b=1/3$) the
stability of critical points are the same as the scalar field[14]
which is no coupling to baryotropic matter. In this case there are
in fact no coupling between $\sigma$ filed and radiation. However,
if the baryotropic fluid is matter($w_b=0$), the situation is
quite different. Since there exists coupling between $\sigma$
field and matter, which is mathematically equivalent to the
coupled quintessence model. It is possible for all the critical
points to be stable but for every value of the parameters
$\lambda$ and $\beta$, there is one and only one critical point
 to be stable; the others are unstable. We plot the phase
 plane for each critical point and the results clearly show the attractor properties of these critical points. With the
constraints from the observation data, we conclude that our
present universe does not reach the ultimate cosmological state
yet and we are all on the way toward final late-time attractor.
Finally we present a numerical computations on the density
parameter $\Omega$ and the decelerating factor $q$. All the
results are consistent with current observation data.
 \section{Acknowledgement}
 \hspace*{15 pt}W.Fang would like to thank Prof. Y.G.Gong for useful comment. We would also thank Prof. L.Amendola for pointing out his work
 on critical points of Coupled quintessence[18]. This work is partly supported by National Natural Science
 Foundation of China(NNSFC) under Grant No.10573012 and No.10575068
and by Shanghai Municipal Science and Technology Commission
No.04dz05905.

{\noindent\Large \bf References} \small{
\begin{description}
\item {[1]}{D.N.Spergel et al., astro-ph/0603449.}
\item {[2]}{P.J.Steinhardt, L.Wang and I.Zlatev, Phys.Rev.D\textbf{59}, 123504(1999)\\
            P.Steinhardt, in $Critical Problems in Physics$,
            edited by V.L.Fitch and D.R.Marlow(Princeton University Press, Princeton, NJ(1997)}
 \item {[3]}{C.Wetterich, Nucl.Phys.B\textbf{302}, 668(1988);\\
            B.Ratra and P.J.E.Peebles, Phys.Rev.D\textbf{37}:3406(1988);\\
            R.R.Caldwell, R.Dave and P.J.Steinhardt, Phys.Rev.Lett\textbf{80}, 1582(1998);\\
            P,J.Steinhardt, L.Wang and I.Zlatev, Phys.Rev.Lett.\textbf{82}, 896(1996);\\
            X.Z.Li,J.G.Hao,and D.J.Liu, Class.Quantum Grav.\textbf{19}, 6049(2002).}
  \item {[4]}{C.Armend\'{a}riz-Pic\'{o}n, V.Mukhanov and P.J.Steinhardt, Phys.Rev.Lett\textbf{85}, 4438(2000);\\
             C.Armend\'{a}riz-Pic\'{o}n, V.Mukhanov and P.J.Steinhardt, Phys.Rev.D\textbf{63}, 103510(2001);\\
             T.Chiba, Phys.Rev.D\textbf{66}, 063514;\\
             T.Chiba, T.Okabe and M.Yamaguchi, Phys.Rev.D\textbf{62}, 023511(2000);\\
             M.Malquarti, E.J.Copeland, A.R.Liddle and M.Trodden, Phys.Rev.D\textbf{67}, 123503(2003);\\
             R.J.Sherrer, Phys.Rev.Lett.\textbf{93)}, 011301(2004);\\
             L.P.Chimento, Phys.Rev.D\textbf{69}, 123517(2004);\\
             A.Melchiorri, L.Mersini, C.J.Odman and M.Trodden, Phys.Rev.D\textbf{68}, 043509(2003)}
 \item {[5]}{B.Feng, M.Z Li, Y.S.Piao and X.M.Zhang, Phys.Lett.B\textbf{634}, 101-105(2006);\\
             Z.K.Guo, Y.S.Piao, X.M.Zhang and Y.Z.Zhang,Phys.Lett.B\textbf{608}, 177-182(2005);\\
             J.Q.Xia, B.Feng and X.M.Zhang, Mod.Phys.Lett.A\textbf{20}, 2409(2005);\\
             H.Wei, R.G.Cai and D.F.Zeng, Class.Quant.Grav.\textbf{22}, 3189-3202(2005);\\
             G.B.Zhao, J.Q.Xia, M.Z.Li, B.Feng and X.M.Zhang, Phys.Rev.D\textbf{72}, 123515(2005);\\
             P.X.Wu and H.W.Yu, Int.J.Mod.Phys.D\textbf{14}, 1873-1882(2005);\\
             R.Lazkoz and G.Le¨®n, astro-ph/0602590.}
\item {[6]}{Q.G.Huang and M.Li, JCAP\textbf{0408}, 013(2004);\\
            M.Ito, Europhys.Lett.\textbf{71}, 712-715(2005);\\
            Ke Ke and M.Li, Phys.Lett.B\textbf{606}, 173-176(2005);\\
            Q.G.Huang and M.Li, JCAP\textbf{0503}, 001(2005);\\
            Y.G.Gong, B.Wang and Y.Z.Zhang, Phys.Rev.D\textbf{72}, 043510(2005);\\
            X.Zhang, Int.J.Mod.Phys. D\textbf{14}, 1597-1606(2005).}
\item {[7]}{L.R.Abramo, F.Finelli and T.S.Pereira, Phys.Rev.D\textbf{70}, 063517(2004);\\
            H.Q.Lu, Int.J.Mod.Phys. D\textbf{14}, 355-362(2005);\\
            M.R.Garousi, M.Sami and S.Tsujikawa, Phys.Rev.D\textbf{71}, 083005(2005);\\
            M.Novello, M.Makler, L.S.Werneck and C.A.Romero, Phys.Rev.D\textbf{71}, 043515(2005);\\
            H.Q.Lu, Z.G.Huang, W.Fang and P.Y.Ji, hep-th/0504038;\\
            A.Fuzfa and J.M.Alimi, Phys.Rev.D\textbf{73}, 023520(2006);\\
            W.Fang, H.Q.Lu, B.Li and K.F.Zhang, hep-th/0512120(Int.J.Mod.Phys.D, in press);\\
            W.Fang, H.Q.Lu, Z.G.Huang and K.F.Zhang, Int.J.Mod.Phys.D\textbf{15}, 199(2006)(hep-th/0409080)}
\item {8.} {R.R.Caldwell, Phys.Lett.B\textbf{545}, 23(2002);\\
             A.Melchiorri, astro-ph/0406652;\\
             V.Faraoni, Int.J.Mod.Phys.D\textbf{11}, 471(2002);\\
             S.Nojiri and S.D.Odintsov, hep-th/0304131; hep-th/0306212;\\
             E.schulz and M.White, Phys.Rev.D\textbf{64}, 043514(2001);\\
             T.Stachowiak and Szydllowski, hep-th/0307128;\\
             G.W.Gibbons, hep-th/0302199;\\
             A.Feinstein and S.Jhingan, hep-th/0304069;\\
             M.Sami and A.Toporensky, Mod.Phys.Lett.A\textbf{19}, 1509(2004).}
 \item {[9]}{A.Lue, R.Scoccimarro and G.Starkman, Phys.Rev.D\textbf{69}, 044005 (2004);\\
             Shin'ichi Nojiri and S.D.Odintsov, Phys.Rev.D\textbf{68}, 123512(2003);\\
             X.H.Meng and P.Wang, Phys.Lett.B\textbf{584}, 1-7 (2004);\\
             Shin'ichi Nojiri and S.D.Odintsov, Mod.Phys.Lett.A\textbf{19}, 627-638(2004);\\
             S.M.Carroll et al., Phys.Rev.D\textbf{71}, 063513(2005);\\
             V.Faraoni, Phys.Rev. D\textbf{72}, 061501(2005);\\
             K.Koyama, JCAP \textbf{0603}, 017(2006).}
 \item {[10]}{B.Gumjudpai, T.Naskar, M.Sami and S.Tsujikawa, hep-th/0502191;\\
              E.J.Copeland, M.Sami and S.Tsujikawa, hep-th/0603057;\\
              R.Bean and J.Magueijo, Phys.Lett.B\textbf{517}:177-183(2001);\\
              T.Damour, F.Piazza and G.Veneziano, Phys.Rev.Lett.\textbf{89}, 081601(2002);\\
              M.Susperregi, Phys.Rev.D\textbf{68}, 123509(2003);\\
              F.Piazza and S.Tsujikawa, JCAP\textbf{0407}:004(2004);\\
              B.Boisseau, G.Esposito-Farese, D.Polarski and A.A.Starobinsky, Phys.Rev.Lett.\textbf{85}, 2236(2000);\\
              G.Esposito-Farese and D.Polarski, Phys.Rev.D\textbf{63}, 063504(2001);\\
              Z.G.Huang and H.Q.Lu, Int.J.Mod.Phys.D\textbf{15}, 1501(2006).}
\item {[11]}{H.Q.Lu, Z.G.Huang and W.Fang, hep-th/0409309.}
\item {[12]}{H.Q.Lu and K.S.Cheng, Astrophysics and Space Science\textbf{235}, 207(1996); \\
             Y.G.Gong, gr-qc/9809015.}
\item {[13]}{Z.G.Huang, H.Q.Lu and W.Fang, Class.Quant.Grav\textbf{23}, 6215(2006)(hep-th/0604160)}
\item {[14]}{E.J.Copeland, A.R.Liddle and D.Wands, Phys.Rev.D\textbf{57}:4686(1998).}
\item {[15]}{Y.G.Gong et al, Phys.Lett.B\textbf{636}, 286-292(2006);\\
             J.J.Halliwell, Phys.Lett.B\textbf{185}, 341(1987);\\
             A.B.Burd and J.D.Barrow, Nucl.Phys.B\textbf{308}, 929(1988);\\
             A.A.Coley, J.Ibanez and R.J.van den Hoogen, J. Math.Phys.\textbf{38}, 5256(1997);\\
             F.Finelli and R.Brandenberger, Phys.Rev.D\textbf{65}, 103522(2002);\\
             I.P.C.Heard and D.Wands, Class.Quant.Grav.\textbf{19}, 5435-5448 (2002);\\
             E.Elizalde, S.Nojiri and S.D.Odintsov, Phys.Rev.D\textbf{70}, 043539(2004);\\
             N.J.Nunes and D.F.Mota, Mon.Not.Roy.Astron.Soc.\textbf{368}, 751-758 (2006);\\
             J.Hartong, A.Ploegh, T.V.Riet and D.B.Westra, gr-qc/0602077;\\
             C.Wetterich, Nucl.Phys.B\textbf{302}, 668(1988);\\
             D.Wands, E.J.Copeland and A.R.Liddle, Ann.(N.Y.)Acad.Sci.\textbf{688}, 647(1993).}
\item {[16]}{Y.M.Cho,Phys.Rev.Lett\textbf{ 68}, 3133(1992);\\
             T.Damour and K.Nordtvedt,Phys.Rev.D\textbf{48}, 3436(1993).}
\item {[17]}{T.Damour and K.Nordtvedt, Phys.Rev.Lett\textbf{70}, 2217(1993);\\
             R.D.Reasenberg et al., Astrophys.J.\textbf{234L}, 219(1979);\\
             B.Bertotti, L.Iess and P.Tortora, Nature\textbf{425}, 374(2003).}
\item {[18]}{L.Amendola, Phys.Rev.D\textbf{62}, 043511(2000); Phys.Rev.D\textbf{60}, 043501(1999)}
\end{description}}
\end{document}